\documentclass[prl,twocolumn,showpacs,preprintnumbers,superscriptaddress,amsmath,amssymb]{revtex4}

\usepackage{color,graphicx,amsmath,upgreek}
\begin{document}
\title{An integrated source of spectrally filtered correlated photons for large scale quantum photonic systems}

\author{Nicholas C. Harris$^*$}
\affiliation{Department of Electrical Engineering and Computer Science,
Massachusetts Institute of Technology, 77 Massachusetts Avenue, Cambridge, MA 02139, USA}
\author{Davide Grassani$^*$}
\affiliation{Dipartimento di Fisica, Universit\`{a} degli Studi di Pavia, via Bassi 6, 27100 Pavia, Italy}
\author{Angelica Simbula}
\affiliation{Dipartimento di Fisica, Universit\`{a} degli Studi di Pavia, via Bassi 6, 27100 Pavia, Italy}
\author{Mihir Pant}
\affiliation{Department of Electrical Engineering and Computer Science,
Massachusetts Institute of Technology, 77 Massachusetts Avenue, Cambridge, MA 02139, USA}
\author{Matteo Galli}
\affiliation{Dipartimento di Fisica, Universit\`{a} degli Studi di Pavia, via Bassi 6, 27100 Pavia, Italy}
\author{Tom Baehr-Jones}
\affiliation{East West Photonics Pte Ltd, 261 Waterloo Street 03-32, Waterloo Centre, 180261 Singapore}
\author{Michael Hochberg}
\affiliation{East West Photonics Pte Ltd, 261 Waterloo Street 03-32, Waterloo Centre, 180261 Singapore}
\author{Dirk Englund}
\affiliation{Department of Electrical Engineering and Computer Science,
Massachusetts Institute of Technology, 77 Massachusetts Avenue, Cambridge, MA 02139, USA}
\author{Daniele Bajoni}
\affiliation{Dipartimento di Ingegneria Industriale e dell'Informazione, Universit\`{a} degli Studi di Pavia, via Ferrata 1, 27100 Pavia, Italy}
\author{Christophe Galland}
\affiliation{Ecole Polytechnique F\'ed\'erale de Lausanne, 1015 Lausanne, Switzerland\\ $^*$ These authors contributed equally to this work}

\date{\today}

\begin{abstract}

We demonstrate the generation of quantum-correlated photon-pairs combined with the spectral filtering of the pump field by more than 95~dB using Bragg reflectors and electrically tunable ring resonators. Moreover, we perform demultiplexing and routing of signal and idler photons after transferring them via a fiber to a second identical chip. Non-classical two-photon temporal correlations with a coincidence-to-accidental ratio of 50 are measured without further off-chip filtering. Our system, fabricated with high yield and reproducibility in a CMOS process, paves the way toward truly large-scale quantum photonic circuits by allowing sources and detectors of single photons to be integrated on the same chip.

\end{abstract}


\maketitle

\section{Introduction}

\begin{figure*}
\begin{center}
\includegraphics[width= 14cm]{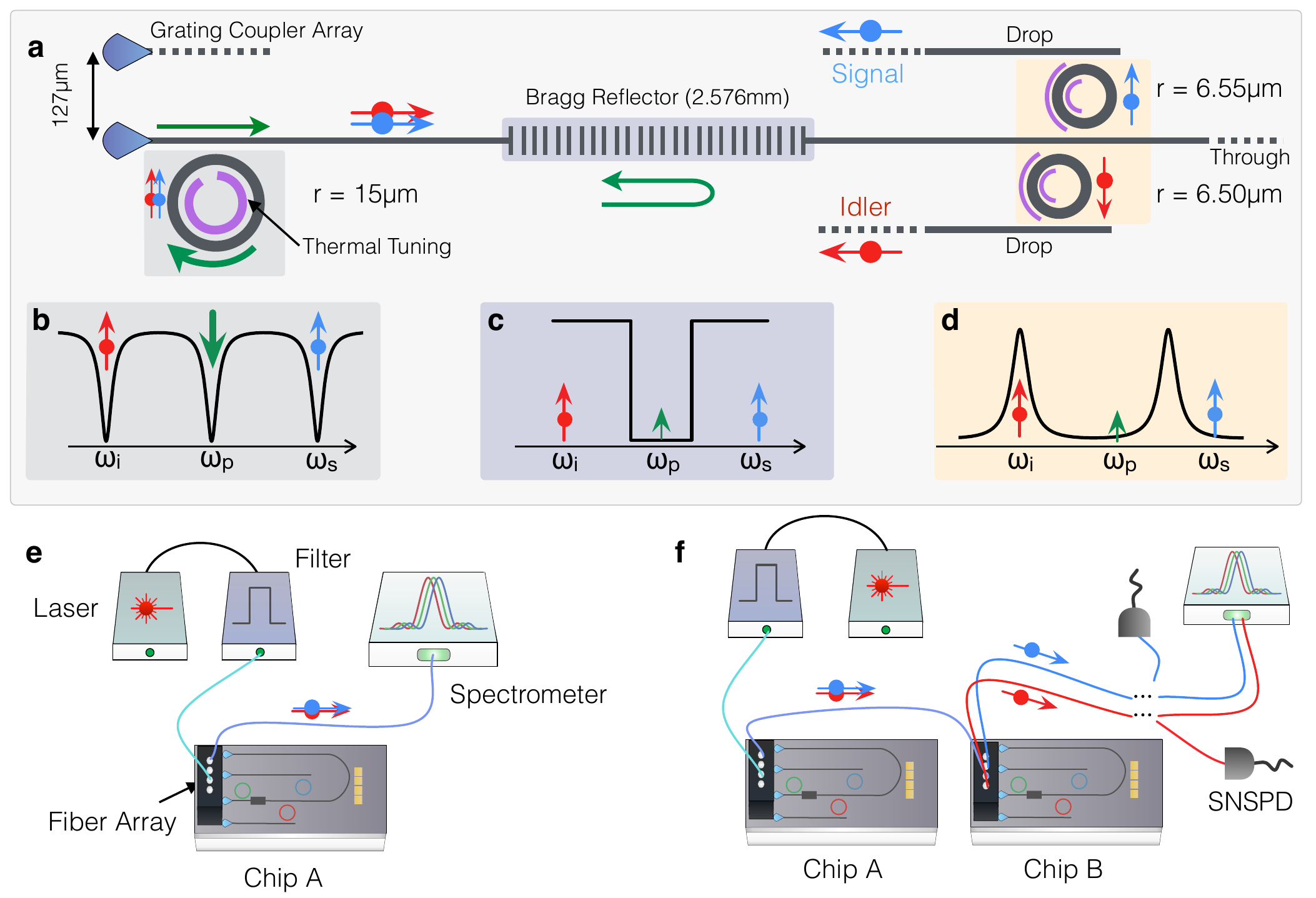}
\caption{(color online)
(a) Schematic layout of the photonic integrated circuit composed of a high-$Q$ thermally tunable ring for efficient pair generation by spontaneous four-wave mixing, followed by a distributed Bragg reflector (DBR) for pump rejection and the add-drop ring resonator filters for demultiplexing of signal and idler photons. Convenient optical coupling to a single-mode polarization-maintaining fiber array is achieved via focusing grating couplers separated by a 127~$\upmu$m pitch.
(b) Schematic transmission spectrum of the first ring around the pump wavelength $\omega_p$. When one of the ring resonances is tuned to the laser at $\omega_p$ signal and idler photons are produced in correlated pairs at neighboring resonance wavelengths $\omega_s$ and $\omega_i$, respectively (pairs are also generated at wavelengths spaced by multiple free-spectral-ranges).
(c) Schematic transmission spectrum of the DBR with the stop band overlapping with the pump wavelength $\omega_p$.
(d) Add-drop filter spectrum tuned to route idler photons to the drop port.
(e) First experimental setup: single-chip pump rejection. The add-drop rings are both tuned on resonance with the pump. Light is collected from the common through port.
(f) Second experimental setup (see Supplemental Fig.~4 for a photograph): Correlated photon pairs  generated in Chip A are sent via a fiber to Chip B where further pump rejection and signal/idler demultiplexing are performed before spectral characterization or coincidence measurements with off-chip superconducting nanowire single photon detectors (SNSPD).
}
\label{intro}
\end{center}
\end{figure*}

Integrated photonic circuits have emerged as a promising platform for quantum information processing \cite{Silverstone2014,Collins:2013eu,Tanzilli2012}. The most mature technology for scalable photonic circuits is based on CMOS-compatible materials such as silicon, silicon nitride, and silica. In these platforms, pairs of quantum-correlated photons produced by spontaneous four-wave mixing (sFWM) are an essential resource for quantum technologies, including linear quantum optics experiments \cite{Spring2013,AspuruGuzik:2012hoa,Politi2008,Peruzzo2010} and quantum sensing \cite{Matthews:2011bq}.

Photon-pair generation via sFWM involves pumping a non-linear medium (such as silicon) with a strong laser field, which must subsequently be filtered to isolate the much weaker signal and idler fields and perform single-photon detection. However, for scalable integration into quantum optics circuits, two challenges must be solved: the on-chip rejection of the pump field and spectral demultiplexing. In particular, and despite very recent attempts \cite{Ong2014,Matsuda2014}, the filtering of the pump field, which requires an extinction on the order of $\sim$100~dB, has never been achieved without using off-chip filters.

Here we demonstrate the generation of time-correlated photon-pairs in a silicon microring with on-chip rejection of the pump field. We also perform chip-to-chip transfer of the photon-pairs via an optical fiber, elimination of the pump field, and signal / idler  demultiplexing before off-chip single-photon detection. A set of four single photon sources with pump rejection and demultiplexing circuitry in a 3.3~mm$^2$ area is fabricated in a standard CMOS silicon photonics process featuring high yield and reproducibility.

The source components are shown in Fig. \ref{intro}. Photon pair generation takes place in an electrically-tunable ring resonator \cite{YarivYeh,Little1998,Bogaerts2012,Almeida2004,Xu2006} via sFWM \cite{Turner2008,Clemmen2009,Azzini2012OL,Azzini2012OE,Engin2013,Reimer2014} (Fig.~\ref{intro}(a-b)). A first stage of pump rejection by more 60 dB is achieved in a notch filter implemented by a 2.576 mm long distributed Bragg reflector (DBR) consisting of a corrugated waveguide (Fig.~\ref{intro}(c)) \cite{Wang2012}. In a first experiment performed on a single chip (Chip A, Fig.~\ref{intro}(e)), two thermally tunable add-drop ring resonators are used to filter the remaining pump light, thereby totaling close to 100~dB extinction ratio. At the output of Chip A, the pump power is less than 1/10$^\text{th}$ of the combined signal and idler fields.

In a second experiment, we transfer the photon pairs from Chip A via a silica fiber connection to a second identical system (Chip B, Fig.~\ref{intro}(f)) in which we filter any residual pump with the DBR and use the add-drop rings to demultiplex the signal and idler photons (Fig.~\ref{intro}(d)). These photon pairs are then routed to different waveguides before detection with off-chip superconducting nanowire single photon detectors \cite{Najafi:2014td, Sahin:2013eb,Sprengers:2011er} (SNSPDs) and time-correlation measurements. The coincidence-to-accidental ratio exceeds $50$ for a continuous pump power of approximately 0.3 mW, confirming successful on-chip full suppression of the pump light and demultiplexing of signal and idler photons.

Together with previous realizations of (i) integrated laser sources \cite{Fang2006,Tanaka2012,Camacho2012,Keyvaninia2013,Creazzo2013,Lee2014,Shuyu2014}, (ii) on-chip sources of quantum states of light \cite{Clemmen2009,Azzini2012OE,Engin2013,Reimer2014,Sharping2006,Lanco2006,Takesue2007,Harada2008,Xiong2011,Matsuda2012,Davanco2012,Olislager2013,Orieux2013,Takesue2014}   (iii) on-chip quantum state manipulation \cite{Politi2008,Politi2009,Matthews2009,Crespi2011,Shadbolt2012,Bonneau2012,Metcalf2013,Silverstone2014}, and (iv) on-chip single photon detection \cite{Najafi:2014td,Schuck:2013ff,Hadfield:2009ug,Pernice:2012uj}, our work addresses the challenge of integrating single photon sources and detectors on the same chip. Our results highlight the promises of CMOS photonics for emerging quantum technologies such as quantum key distribution \cite{Bennett:1984wv,Gisin2002,Ekert1991,Branciard2005,Renner2005,Brassard2000,Curty2004,Deutsch1996,Marcikic2004,Bennett1994,Mower2013}, quantum simulations and random walks \cite{Metcalf2013,Broome2013,Childs2013,Crespi2013,Tillmann2013,Spring2013,Spagnolo2014} and possibly quantum computation \cite{NielsenChuang,Gottesman1999,Knill2001,Jozsa2003,Gao2010,Ladd2010}.

\section{System Design}

\begin{figure*}
\begin{center}
\includegraphics[width= 14cm]{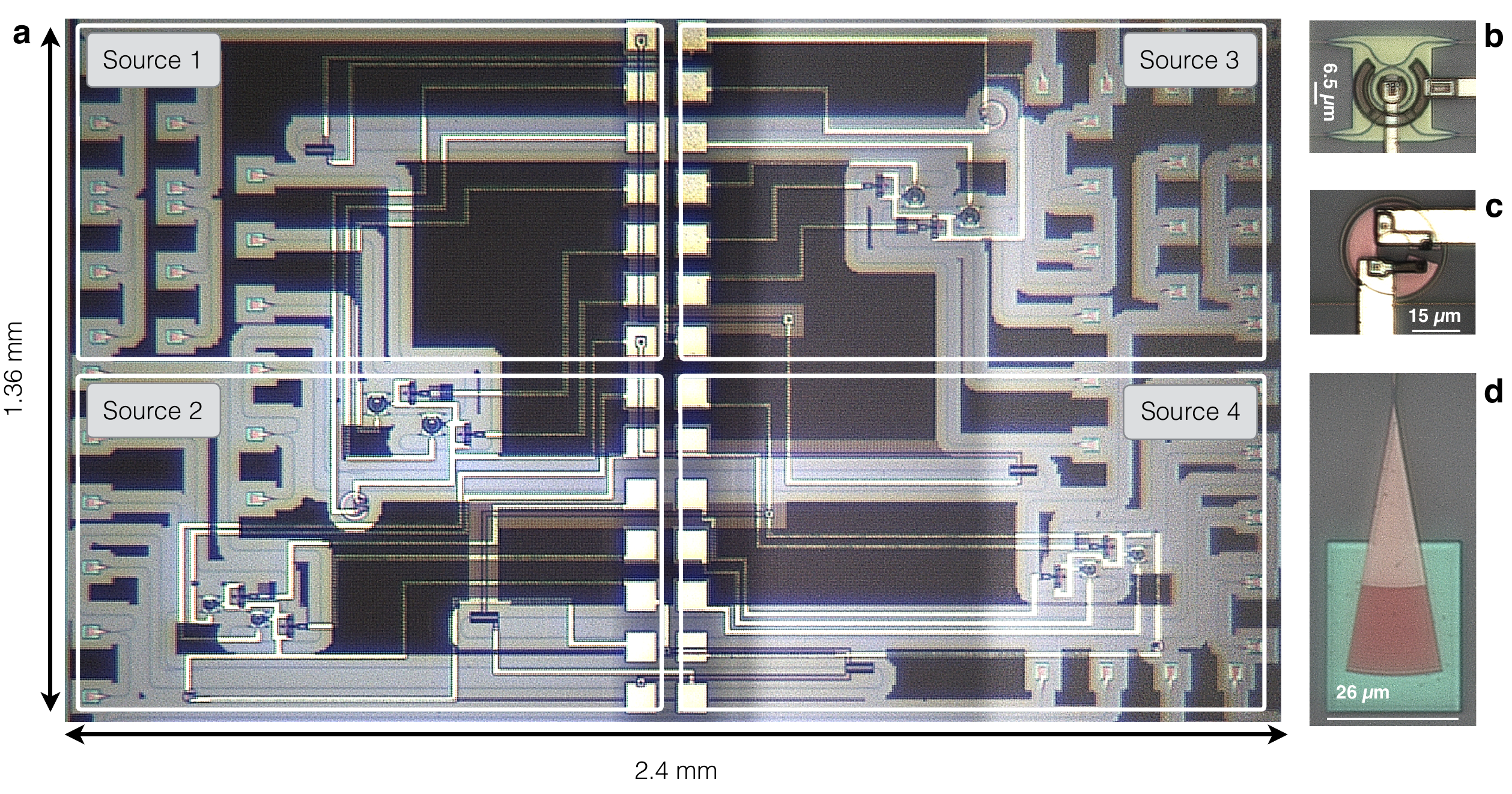}
\caption{(color online)
(a) Optical micrograph of a chip. Four sources are visible with electrical contact pads running along the vertical mid-line.
(b) Close view of an electrically tunable add-drop ring resonator used for either pump filtering or signal and idler demultiplexing.
(c) Close view of an electrically tunable photon-pair generation ring.
(d) Grating coupler used to couple (collect) light to (from) the system.
}
\label{systems}
\end{center}
\end{figure*}

(See Supplemental Material for further details)

\emph{Pair generation---} As a source of photon pairs, we employ a silicon ring resonator evanescently coupled to the input waveguide. Field enhancement inside the resonator results in an increased generation of correlated photon pairs (with respect to non-cavity based generation methods) \cite{Clemmen2009,Azzini2012OE,Engin2013,Helt2010} when the pump ($\omega_p$), signal ($\omega_s$) and idler ($\omega_i$) frequencies match a triplet of ring resonances fulfilling $\omega_i+\omega_s=2\omega_p $. For a given pump power, the flux of photon pairs produced by sFWM is proportional to the ratio $\frac{Q^3}{R^2}$, where $Q$ is the quality factor and $R$ the radius of the ring \cite{Azzini2012OL,Helt2010}. Based on these considerations, we designed the pair-generation ring  (Fig.~\ref{intro}(b)) with a radius of 15 $\upmu$m, leading to an expected free spectral range of 6~nm. To allow for spectral tunability without compromising on quality factor, we added an inner semi-ring of doped silicon to form a resistive heater. 

\begin{figure*}
\begin{center}
\includegraphics[width= \textwidth]{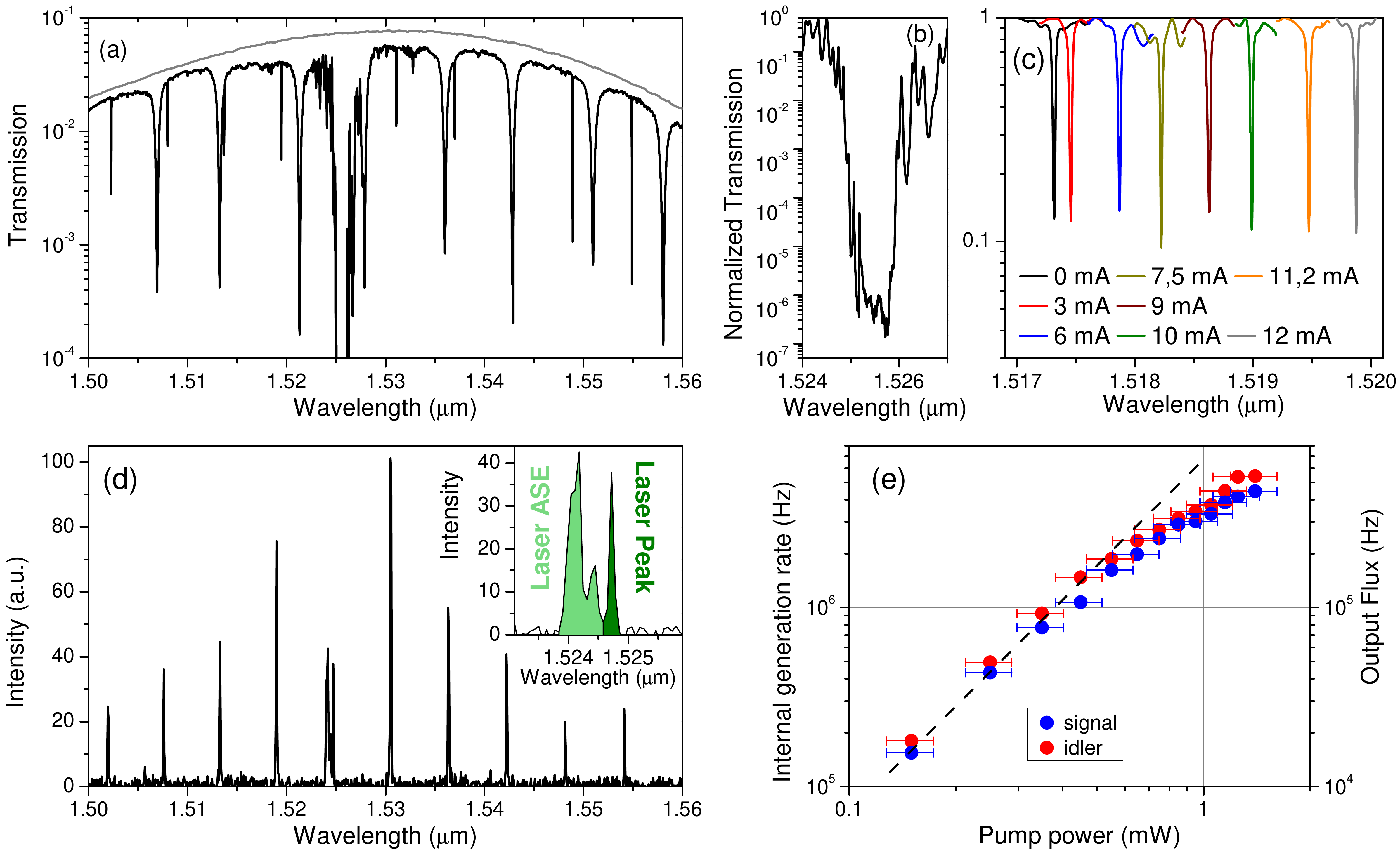}
\caption{(color online)
(a) Logarithmic plot of the transmission spectrum measured at the through port of Chip A after propagation through the complete system. The upper curve is the transmission of a grating coupler loop.
(b) Higher resolution spectrum around the DBR stop band.
(c) Spectrum of the pair-generation ring around the signal wavelength under increasing current driving the resistive heater. 11.2 mA is the value used in the coincidence measurements presented in Fig.~\ref{Fig:twochip}.
(e) Dependence of the sFWM generation rate for the signal (blue) and idler (red) photons as a function of on-chip pump power (dashed line is a guide to the eye proportional to $P_{pump}^2$). Right-hand scale is the off-chip count rate after correction for the CCD efficiency. Left-hand scale is the estimated internal generation rate inside the ring resonator.}
\label{Fig:single}
\end{center}
\end{figure*}

\emph{Pump rejection---} We expect the ratio between the pump power inside the ring resonator and the generated signal and idler beams to be of the order of $10^9$ to $10^{10}$, and thus pump rejection of $\simeq$ 100~dB is required. Achieving pump rejection levels on this order constitutes the most demanding requirement for the system. The center reflection wavelength of the DBR notch filter \cite{Wang2012} is designed to be $\lambda_0 = 2 n_\text{eff}\Lambda\sim1536$~nm, as defined by the Bragg period $\Lambda=320$~nm and the effective index $n_\text{eff}\sim 2.4$. We designed the width of the reflection band to be smaller than the generation ring free spectral range (see Fig.~\ref{intro}(c)), but large enough to avoid the need for delicate spectral tuning. This width depends on the refractive index contrast of the grating (given here by the amplitude of the sidewall corrugation). A 60~nm modulation of the waveguide width was found to yield spectral linewidths of 1-2~nm. The transmission in the reflection band drops exponentially with the number of grating periods. While developing the DBR device, we measured extinction ratio of $\sim$20-25~dB for devices with 2000 periods---prompting the choice of 8000 periods for the system presented here. To reduce the total footprint, we introduced a bend halfway in the DBR waveguide, which leads to weak Fabry-Perot effects as can be seen in Fig.~\ref{Fig:single}(b).

\emph{Signal/idler demultiplexing---} The add-drop ring resonators used to filter the remaining pump and/or route signal and idler photons were designed to cause minimal excess loss, to have a free spectral range about 2.5 times larger than the one of the pair-generation ring, and to be thermally tunable by an embedded resistive heater formed by doped silicon regions contacted to the metal interconnect layer. To minimize losses due to free-carrier absorption, we use a low dopant concentration in the waveguide region overlapping with the optical mode. To maximize the collection efficiency in the drop-port (where single photons are routed), we designed the device to be over-coupled.

\emph{Optical in/out coupling---} An array of optimized non-uniform focusing grating couplers \cite{He:2013hx} offer a convenient and efficient (~4 dB insertion loss, see Supplement) way to couple light between a single-mode fiber array and the silicon waveguides. Moreover the large mode field diameter of the grating couplers (10~$\upmu m$) enables stable optical coupling.

\section{Fabrication}
The system was fabricated in a CMOS-compatible foundry service (OpSIS) with 248~nm lithography on an 8" silicon-on-insulator (SOI) wafer with a 220 nm thick epitaxial silicon layer (bulk refractive index $n_{\text{Si}}$ = 3.48 at 1550~nm) on top of 2 $\upmu$m buried oxide and covered by 2 $\upmu$m oxide cladding (bulk index $n_{\text{SiO}_2}$ = 1.46). The grating couplers and rib-waveguides were defined by 60~nm and 130~nm deep anisotropic dry etching, respectively. A final etch step down to the buried oxide was used to pattern the 500~nm wide ridge waveguides designed to be single-mode in the 1500-1600~nm wavelength range. \textit{N}-type doping of the tunable ring resonators was realized by phosphorous implantation on the exposed silicon before oxide cladding. Dopants were activated by a rapid thermal anneal (RTA) at 1030$^\circ$C for 5 seconds. The back-end-of-line consists of two levels of aluminum interconnects contacted by aluminum vias. Further technical details on the process can be found in Ref. \cite{baehr201225}. A micrograph of the chip area containing four similar complete systems is reproduced in Fig.~\ref{systems}.

\section{Single-chip Pump Rejection}

A high-resolution transmission spectrum of the complete system is shown in Fig.~\ref{Fig:single}(a). This spectrum was acquired using a tunable laser and a high dynamic-range power meter. The dip around 1525~nm is due to the DBR and is detailed in the high resolution spectrum of Fig.~\ref{Fig:single}(b)). In control experiments, we observed that some of the laser light is directly coupled from the input to the output fibers -- without passing through the waveguides -- by scattering and reflections in the SiO$_2$ cladding and Si substrate. As a result, the measured rejection of the DBR is reduced with respect to the expected value of 80-100~dB to $\sim$65~dB with an insertion loss of approximately 3 dB. Weak Fabry-Perot resonances are visible near the red edge of the stop band due to the cavity formed by the bend separating the two halves of the DBR. Better rejection could be achieved in a future implementation by having a larger spatial separation between the input and output grating couplers to avoid collecting scattered pump light.

The wider Lorentzian resonances in the spectra are due to the two final add-drop rings, with a quality factor ($Q$) of $\sim 4 \times 10^3$ and an insertion loss of $\sim$1.5 dB. The narrow Lorentzian resonances are due to the photon-pair generation ring, with a $Q$ of $\sim 4\times10^4$ (corresponding to a photon lifetime of about 30 ps). Such high quality factor is of fundamental importance in providing a sufficiently high generation rate at an easily achievable mW pump power level \cite{Azzini2012OL} and for ensuring emission of quantum states with narrow bandwidth \cite{Fujii2007}.

The spectrum is attenuated by $\simeq$ 20\% due to a directional coupler designed for monitoring of the system under operation. The sinusoidal envelope is due to the grating coupler spectral response (measured insertion loss of 5 dB) and depends on the tilt angle of the input and output fibers. The transmission from a grating coupler loop is also shown in Fig.~\ref{Fig:single}(a) for comparison. The fiber array tilt angle was chosen to optimize the coupling around the pump wavelength ($\sim$1525~nm).

The setup for the first pair-generation experiment is shown in Fig.~\ref{intro}(e). The external pump laser is cleaned through a tunable band-pass filter to suppress the sidebands below the expected sFWM yield. It is coupled to Chip A via an 8-port fiber array aligned with the focusing grating couplers \cite{He:2013hx}. The pump laser and the generation ring are tuned in resonance with each other so that their wavelength lie within the DBR stop band (Fig.~\ref{Fig:single}(a)). Inside the ring, signal and idler photons are spontaneously generated in pairs at resonance frequencies symmetrically detuned from the pump. The pump light is first rejected by the DBR (Fig.~\ref{Fig:single}(b)) and then filtered by the two add-drop filters, which are tuned on resonance with the pump wavelength.

The spectrum of the light collected at the common through-port, for a coupled laser power of 1 mW, is plotted in Fig.~\ref{Fig:single}(d). Residual light from the pump is still visible around 1524.7~nm, but it is greatly reduced. We estimate a total extinction ratio of about 95-100~dB. Notice that part of the amplified spontaneous emission (ASE) of the laser is still visible in our spectrum (Fig.~\ref{Fig:single}(d), inset). This could be suppressed by using a filter with narrower bandwidth (instead of 0.15~nm here) to clean the laser line (which could easily be implemented on-chip too).
The sum of the intensities in the first five signal peaks (resp. in the first five idler peaks) is a factor of 8 (resp. 10) larger than the residual laser. Remaining pump photons follow Poissonian statistics and are not correlated in time, therefore they do not affect the CAR significantly. This suppression is already sufficient \cite{Takesue2010} for many quantum optical experiments like heralding of single photons or entanglement generation.

The dependence of signal and idler intensities on the pump power is plotted in Fig.~\ref{Fig:single}(e). Both intensities follow the quadratic dependence on  pump power expected for sFWM, and around 1~mW on-chip pump power saturation due to two photon absorption \cite{Xu2006} becomes visible. To measure the photon rate at the sample output, we have calibrated our CCD camera using a high sensitivity power meter. The internal generation rate, on the order of several MHz, was then estimated by subtracting the losses on the path from the output to the generating ring to the spectrometer.

\section{Chip-to-chip transfer, Demultiplexing and Correlation measurements}

\begin{figure*}
\begin{center}
\includegraphics[width= \textwidth]{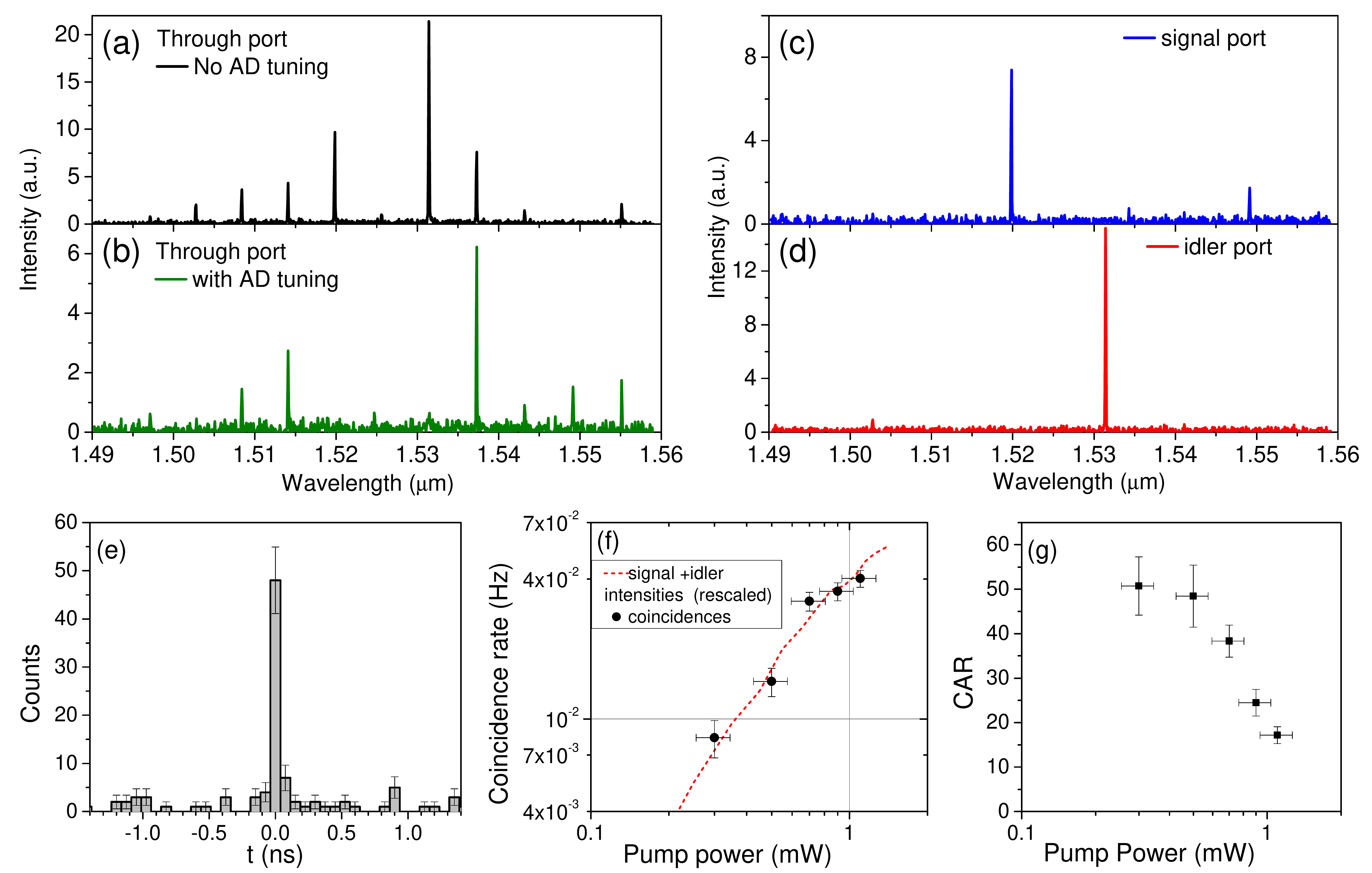}
\caption{(a) sFWM spectra as observed at the through-port of Chip B without tuning the add-drop ring filters (black line) and (b) with add-drop filters tuned to the signal and idler wavelengths (green line).
(c) and (d) show the sFWM spectra as observed at the idler (red) and signal drop ports (blue) of Chip B, respectively, when the add-drop filters are tuned as in (b). In all panels $P_{pump}$= 0.5~mW in Chip A.
(e) Coincidence histogram between signal and idler ports of Chip B for a pump power $P_{pump}$=0.5 mW in Chip A. The total acquisition time was 50~minutes.
(f) Dependence of the coincidence rate on the pump power. The dashed line is proportional to the measured signal and idler combined intensities.
(g) Pump power dependence of the coincidental-to-accidental ratio (CAR).}
\label{Fig:twochip}
\end{center}
\end{figure*}

In a second experiment (Fig.~\ref{intro}(f)), we send the output of Chip A (without tuning the add-drop rings for pump rejection) via a fiber to a second, identical system. In this experiment, the DBR of Chip B is tuned in resonance with the pump by temperature control of the full chip (see Supplemental Fig.~3(d)) and the add-drop filters on Chip B are tuned individually to perform demultiplexing of the signal and idler photons (see Fig.~\ref{intro}(d)), while further eliminating the residual pump photons.

Figs.~\ref{Fig:twochip}(a-b) show the spectrum of the common through-port on Chip B before (a) and after (b) tuning the add-drop rings. It demonstrates (i) the efficient generation of signal and idler photons by sFWM at the resonance wavelengths of the pair-generation ring and (ii) the almost complete rejection of the pump beam (missing central peak around 1525~nm). The pronounced decrease in the intensity of the generated fields moving away from the pump wavelength is due to the spectral response of input and output grating couplers (see Fig.~\ref{Fig:single}(a)).

As desired, the signal and idler emission lines are not visible in the through port after tuning. The corresponding spectra from the signal and idler through-ports are shown in Fig.~\ref{Fig:twochip}(c) and (d), respectively. Each resonance is correctly routed to an individual through-port, and no trace of the pump or of the opposite resonance is visible. Our data indicate the first realization of on-chip sFWM with pump rejection and signal/idler demultiplexing achieved without any off-chip post-filtering elements.

We measure two-photon time-correlations between the two drop-port outputs of Chip B using two SNSPDs with 10\% and 5\% quantum efficiency, without external filters. The bias current was set to have dark count rates of the order of 300~Hz. An example of coincidence measurement is shown in Fig.~\ref{Fig:twochip}(e). A clear coincidence peak is visible at zero time delay. Accidental events outside the central peak are mainly caused by coincidences between dark counts and the detection of only one photon in a pair.

The dependence of the coincidence rate after Chip B on the pump power inside Chip A is shown in Fig.~\ref{Fig:twochip}(f). The dependence closely follows the combined signal and idler emission rates, reported in the figure after rescaling for clarity. The total losses on the coincidences are given from the product of the total losses on both channels, from the generating ring to the detectors, amounting to a total of 68 dB (see discussion in Supplemental Material). Finally, in Fig.~\ref{Fig:twochip}(g) we show the coincidence-to-accidental ratio \cite{Takesue2010,Azzini2012OE,Husko2013} as a function of the pump power in Chip A. A value of 50$\pm6$ is achieved under 0.3~mW pump power. This is far above the classical limit of 2, and allows for high fidelity preparation of entangled photon pairs or heralded single photons.

\section{Discussion and outlook}

Quantum information processing based on linear-optics quantum computing (LOQC) will require efficient single-photon sources and detectors as well as feed-forward operations \cite{Knill2001,Varnava:2008bd}. The potential for multiplexing the emission of parametric single-photon sources \cite{PhysRevA.66.053805,2011.PRA.Mower.AMPP} could enable high-efficiency state preparation for quantum computation. The system presented here permits the integration of single photon sources and detectors on a single chip, providing a means of achieving large-scale systems for LOQC.

Boson sampling has received significant attention \cite{Broome2013,Spring2013,Tillmann2013} for its promise to demonstrate the first example of quantum speedup over the fastest known classical algorithm. Leveraging the reproducibility and high yield of integrated photonics, our system could be tiled to enable boson sampling with many sources. The signal photon could be used as a herald for the idler photon (or vice versa) in a scheme such as scattershot boson sampling \cite{Lund:2013wt} where the measurements are post-selected based on the number of photons that entered the quantum random walk simultaneously. The dimension of the input state can thereby be scaled up in a probabilistic scheme.

Finally, it has been shown that the sFWM sources used here produce time-energy entangled photon-pairs \cite{Grassani2014}. In light of this, our  architecture could enable fully-integrated quantum key distribution (QKD) emitters and receivers based on time-energy entanglement protocols \cite{Zhang:2014gt,Mower2013}. The geometry of nanoscale silicon waveguides can be engineered to achieve both normal and anomalous dispersion, as necessary to realize this protocol.

\section{Conclusion}

We have demonstrated, for the first time in a monolithic, tunable silicon photonic chip, the generation of quantum-correlated photon pairs in an electrically tunable ring resonator and the rejection of the pump field by more than 95 dB using a Bragg reflector and tunable ring filters---enabling the integration of single-photon sources and single-photon detectors on chip \cite{Najafi:2014td,Schuck:2013ff}. Moreover, we achieved complete pump rejection and spatial demultiplexing of signal and idler photons by employing two identical chips connected by an optical fiber link. With no additional off-chip filtering, we measured signal-idler temporal correlations with a coincidence-to-accidental ratio of 50 at the output of the demultiplexing chip, confirming their non-classical nature.  Each source, whose total footprint is less than 1~mm$^2$, is fabricated using a conventional CMOS-compatible photonic process featuring high yield, reproducibility, dense integration and scalability. By eliminating the need for the last off-chip components, our result opens new possibilities for large-scale quantum photonic systems with on-chip single- and entangled-photon sources.

\section{Acknowledgments}
N.H. acknowledges that this material is based upon work supported by the National Science Foundation Graduate Research Fellowship under Grant No. 1122374. This work was supported in part under a grant from the Air Force Research Laboratory (AFRL/RITA), Grant FA8750-14-2-0120.  Any opinions, findings and conclusions or recommendations expressed in this material are those of the author(s) and do not necessarily reflect the views of the AFRL. The authors are also grateful for the support of Portage Bay Photonics and of Gernot Pomrenke of AFOSR, who has supported the broader research program of OpSIS over multiple programs.
D.G., D.B. and M.G. acknowledge support from MIUR through the FIRB ``Futuro in Ricerca'' project RBFR08XMVY, from the foundation Alma Mater Ticinensis and by
Fondazione Cariplo through project 2010-0523  Nanophotonics for
thin-film photovoltaics. C.G. is supported by an \textit{Ambizione} Fellowship from the Swiss National Science Foundation. C.G. thanks S\'ebastien Tanzilli for valuable comments and discussions.

\bibliographystyle{apsrev-title}

\bibliography{citations}

\end{document}